\documentclass{article}
\usepackage{color}
 \usepackage{amsbsy}
  \usepackage{bm}
 \usepackage{amsfonts}
\usepackage{amssymb}
\usepackage{amsmath}
\usepackage{graphics}
\usepackage{graphicx}
\begin{document}
\title{Influence of hyperon-hyperon interaction on the properties of
neutron stars}
\author{R. M. Aguirre
\footnote{Email address: aguirre@fisica.unlp.edu.ar}}
\date{\it{Departamento de Fisica, Facultad de Ciencias Exactas}, \\\it{Universidad Nacional de La Plata,} \\
\it{and IFLP, UNLP-CONICET, C.C. 67 (1900) La Plata, Argentina.}}


\maketitle

\begin{abstract}
The properties of neutron stars are studied in a composite model
of the strong interaction. In the regime of low to medium baryonic
densities a covariant hadronic model is adopted which includes an
exclusive channel for the hyperon-hyperon interaction mediated by
hidden strangeness mesons, which in turn couple to other mesons
through polynomial vertices. The new coupling constants are
subject to phenomenological constraints. The presence of free
quarks in the core of the star is considered by using the
Nambu-Jona Lasinio model supplemented with a vector interaction.
The deconfinement process is described by a continuous coexistence
of phases. Several structure parameters of neutron stars, such as
mass-radius relation, moment of inertia, tidal deformability, and
the propagation of nonradial f and g-modes within the relativistic
Cowling approximation are studied. The predictions of the model
are in good agreement with recent observational data, in
particular the maximum inertial mass is greater than the
observational lower limit of two solar masses.

 \noindent
\\

\end{abstract}

\section{Introduction}
The astronomical observation of compact stars has became the most
promising source of evidence for the study of matter at extreme
densities and low temperatures. It is expected that the different
manifestations of the interacting matter  will leave its imprint
in specific aspects of the structure and dynamics of these stars.
From the theoretical point of view, the study of neutron stars is
an excellent opportunity to try a unified theoretical description
covering the many facets of the strong interaction in combination
with gravitation, and to contrast predictions with empirical
results.\\
The field  has been a subject of permanent growing and interest,
and the accelerated advances in the observational techniques has
concentrated multiplied efforts in recent years. Thus a large
amount of data has been acquired and is able for analysis. With
the prospect of new and more precise measurements under an
improved technology, it is expected that this trend will continue
in the near future.

Some properties of the stellar structure are more transparent for
understanding the microscopic behavior of the matter that compose
it, and are taken as true guidelines to extract local information
from the global experimental evidence. A few of them are
highlighted in
the following paragraphs.\\
The lowest upper bound for the inertial mass of a neutron star has
been established around   $M_\text{max}\simeq 2\, M_\odot$
according to recent evidence \cite{DEMOREST,ROMANI,FONSECA}.
Whereas it is not clear yet if the report \cite{ABBOTT0} of a
compact object with a mass  $M/M_\odot=2.5 \,-\, 2.67 $
corresponds to a black hole or to a neutron star. In the last case
a serious questioning is possed against most of the descriptions
of the strong interacting matter used at present \cite{FATTOYEV0}.
In fact only a few microscopic models are able to reach such large
value for the inertial mass. They commonly use the conventional
nuclear degrees of freedom from the crust to the core of the star.
The equation of state (EoS) of the dense matter becomes softer as
other energetically favorable configurations are considered. The
rise of the hyperon population \cite{BEDNAREK,LIM hyp,BURGIO}, the
emergence of resonances such as the $\Delta$ isobar
\cite{CAI,SAHOO,SUN,RIBES,SEN,MARCZENKO}, the meson condensates
\cite{SHAO,LIM,THAPA}, the deconfinement phase transition
\cite{PAOLI}, or the mixing with dark matter
\cite{ELLIS,DAS,LEUNG}, are examples where the maximum mass is
reduced due to the softening of the EoS.\\
The mass-radius relation has been inferred for the object PSR
J0030+0451 \cite{RILEY} obtaining $M=1.34^{+0.15}_{-0.16}
M_\odot$, $R=12.71^{+1.14}_{-1.19}$ km. For the case PSR
J0740+6620 the results are $M=2.072^{+0.067}_{-0.066} M_\odot$,
and $R=12.39^{+1.30}_{-0.98}$ km \cite{RILEY2}. Combining these
and other experimental evidence the radius of the star with
$M/M_\odot=1.4$ has been estimated within the range $11.4$ km $< R
< 13.1$ km  \cite{RAAIJMAKERS}. This result rules out some
commonly used models, such as MS0 \cite{MUELLER} and NL3
\cite{LALA}, which employ exclusively the low density nuclear
degrees of freedom to reach the condition $M_\text{max}\geq 2.5\,
M_\odot$. Although they would be admissible when combined with
dark matter models to describe homogeneously admixed stars
\cite{DASMALIK,DASKUMAR}.
\\
The moment of inertia encodes valuable information about the
organization of matter inside the star and can be related to the
formation and evolution of double systems of compact objects
\cite{NEWTON}. There is great expectation for the measurement of
the moment of inertia of one of the components of the pulsar
binary system PSR J0737-3039 \cite{LYNE}, which is particularly
favorable for this purpose. Different procedures has been proposed
to distinguish the underlying EoS, such as comparison with the
prediction of specific models
\cite{MORRISON,LATTIMER,RAITHEL,GREIF}, use of empirical universal
relations \cite{LANDRY}, and statistical
analysis \cite{LIMHOLT}.\\
Closely related to the measurement of the star masses and moments of
inertia is the mutual deformation of a double star system due to
gravitation. It has been proposed that tidal deformability, i.e. the
quotient of the quadrupole deformation to the perturbing tidal
field, is the relevant quantity to describe the gravitational wave
phase emitted in the early steps of the collapse of a binary system
\cite{HINDERER}. Thus in the event GW170817 the masses of the pair
has been determined either as $1.36 < m_1/M_\odot < 1.60$ and $1.16
< m_2/M_\odot < 1.37$ for the low spin regime, or $1.36 <
m_1/M_\odot < 2.26$ and $0.86 < m_2/M_\odot < 1.36$ for higher
angular momentum \cite{ABBOTT1}. In addition an upper bound for the
tidal deformability $\Lambda_{1.4}\leq 970 \;(800)$ of a neutron
star with mass $1.4 M_\odot$ was established for each of the regimes
just mentioned. Further refinements \cite{ABBOT2} obtained the
preferable values $1.36 < m_1/M_\odot < 1.62$, $1.15
< m_2/M_\odot < 1.36$, and $\Lambda_{1.4}=190^{+390}_{-120}$.\\
The propagation of nonradial oscillations inside a compact star is
a longstanding issue \cite{CAMPOLATARO,McDERMOTT}, which has
gained renewed interest because of its relation with the
gravitational waves and the recent first detection of such waves
coming from the collapse of a binary system of neutron stars
\cite{ABBOTT1,ABBOT2,ABBOT3}. A class of low frequency
oscillations, known as f and g-modes, have received particular
attention because their $l=2$ multipolar component can couple with
enough intensity to presently observed gravitational waves.
Intensive work has been devoted to establish explicit relations
between the frequency of different modes of pulsation and other
properties of the star \cite{ANDERSSON,LAU,WEN,SOTANIk,CHIRENTI}
or even with the EoS of the dense matter
\cite{WEN,SOTANIy,BLAZQUEZ,SALINAS,PRADHAN,JAIKUMAR,CONSTANTINOU,DASKUMAR2,KUNJIPURAYIL,ZHAO}.\\
The search for such kind of relations has been extended to other
observables, with the aim of obtaining model independent
interpretations of the observational data. The relations between
the moment of inertia and the quadrupole moment in terms of the
tidal Love number for slowly rotating stars are paradigmatic
\cite{YAGI,KUMAR,LANDRY}.

The description of the neutron stars based on microscopic models
of the strong interaction still suffers from important
uncertainties. The contrast between the recently obtained
observational data  and the predictions of a high variety of
hadronic models
\cite{TAKAMI,WEIH,NANDI,LOURENCO,SEDRAKIAN,TRAVERSI,GOMES,HAN,DEXHEIMER,LANDRY2,HOLZ,ESSICK,FATTOYEV,BLACKER,
LAU2,LEGRED,PANG,MOST,FERREIRA,SHANGGUAN,AGUIRRE} is a source for
fix missing information. Many of these studies have assumed that
matter is composed only by protons, neutrons and leptons. Hence
the crucial requisite of neutron stars masses of at least $M
\simeq 2 M_\odot$ is guaranteed \cite{LOURENCO}. A smaller number
of investigations include effects of the hyperons
\cite{SEDRAKIAN,TRAVERSI,GOMES,HAN,DEXHEIMER,LANDRY2,HOLZ,BLACKER,MOST,AGUIRRE}.
Furthermore, the possibility of a deconfinement phase transition
taking place through different realizations has been studied
\cite{GOMES,HAN,PANG}. First order transitions with discontinuous
EoS have received special attention \cite{WEIH,BLACKER,LAU2,MOST}
because their effects  are more evident and would be detectable by
the post-merger gravitational wave \cite{BLACKER}. However the
study of the observational data obtained so far is not conclusive.
For instance, the analysis of the GW170817 event in \cite{HOLZ}
varies according to the amount of a priori information deposited
on the sampling of EoSs. The probability for a free quark phase is
$56\%$ against $44\%$ for a pure hadronic phase in one case, but
an inverted $36\%$ against $64\%$ corresponds to the less informed
sample. The analysis made in \cite{PANG} including data from
GW170817 and GW190425 events, does not find evidence of a strong
phase transition.

The present work is mainly devoted to study the hyperon effects on
the structure of neutron stars and on the pulsation modes
propagating in their interiors. So it can be considered as the
continuation of the previous study \cite{AGUIRRE}.\\
It is well known that neutron stars can support a
stable hyperon population at densities well above the normal
nuclear density, producing an energetically favorable state. At
such densities the use of nuclear potentials which instantaneously
propagates the strong interaction is not suitable. For this reason
a covariant model of the field theory of
hadrons is used in the mean field approach. \\
The hyperons persist until extremely large densities, where the
hadronic matter eventually undergoes a transition to a deconfined
quark phase. The rise of the hyperon population as well as the
deconfining transition significantly change the composition of the
core of the star, producing a softer EoS. This, in turn, reduce
the maximum mass achievable by the star. Thus, most of the models
considering hyperons do not reach the measured value  $M/M_\odot
\simeq 2$. This situation is known in the literature as the
hyperon puzzle \cite{BURGIO}.\\
In order to examine the ability of models containing hyperons to
adjust recent astrophysical data, a composite description is
proposed here. For extreme densities one can expect a deconfined
quark phase which is treated within the Nambu-Jona Lasinio model
(NJL) with vector interaction. For the hadronic phase, at medium
densities, the hyperon-hyperon interaction is introduced together
with polynomial meson-meson vertices. A continuous transition
between such phases is assumed.

This work is organized as follows, in the next section the general
theoretical description is presented. Section \ref{Sec3} is
devoted to describe the evaluation of some properties of a neutron
star. Specific results are shown and discussed in Sec. \ref{Sec4}
and finally the conclusions are drawn in Sec. \ref{Sec5}.

\section{Theoretical description}\label{Sec2}

For the hadronic phase a composition of models is considered. In
the very low density regime, corresponding to the crust of the
star, the scheme proposed in \cite{BPS} will be used, as discussed
below. It takes account of the nucleation of protons and neutrons
in equilibrium, surrounded by a sea of electrons.\\
As the density is increased this nuclear system is replaced by a
homogenous fluid of nucleons, according to the description of a
model of the relativistic field theory of hadrons. Within this
same framework, the onset of hyperons becomes energetically
favorable for medium densities and they coexist with the nucleons.\\
Finally, for sufficiently dense matter, the deconfined quarks
emerge as the relevant degrees of freedom. In such case the Nambu-
Jona Lasinio effective model of the strong interaction will be used.\\
In the intermediate density regime the strong interaction is
represented by a model of baryons coupled linearly to mesons, and
the latter exhibit polynomial vertices. The lagrangian density can
be written as
\begin{eqnarray}
{\mathcal L}_H&=&\sum_b \bar{\psi}_b\left(i \not \! \partial -M_b
+g_{\sigma b}\, \sigma+ g_{\xi b} \, \xi+ g_{\delta b}\, \delta -
g_{\omega b} \not \! \omega - g_{\phi b} \not\! \phi - g_{\rho b}
 \not \!
\rho \right) \psi_b  \nonumber\\
&&+ \frac{1}{2}
(\partial^\mu \sigma
\partial_\mu \sigma - m_\sigma^2 \sigma^2)-\frac{A}{3}\, \sigma^3-\frac{B}{4}
\,\sigma^4 + \frac{1}{2} (\partial^\mu \mathbf{\delta} \cdot
\partial_\mu \mathbf{\delta} - m_\delta^2 \delta^2)+ G_{\sigma \delta} \sigma^2 \,\delta^2\nonumber\\
 &&+ \frac{1}{2} (\partial^\mu \xi
\partial_\mu \xi - m_\xi^2 \xi^2)+ G_{\sigma \xi} \sigma^2 \,\xi^2+ G_{\xi \delta} \xi^2 \,\delta^2
-\frac{1}{4} W^{\mu \nu} W_{\mu \nu} + \frac{1}{2} m_\omega^2
\omega^2 \nonumber \\
&&-\frac{1}{4} R^{\mu \nu}\cdot R_{\mu \nu}  + \frac{1}{2} m_r^2 \rho^2+ G_{\omega \rho} \rho^2 \, \omega^2
-\frac{1}{4}
 F^{\mu \nu} F_{\mu \nu} + \frac{1}{2} m_\phi^2\,\phi^2\label{LAGRANGIAN}
\end{eqnarray}

\noindent where the sum runs over the octet of lightest baryons.
In addition to the commonly used $\sigma, \omega, \rho$ mesons,
here the scalar iso-vector field $\delta$,  as well as the hidden
strangeness $\xi, \phi$ mesons are also included. Here $\delta$
and $\rho$ stand for the third isospin component of these meson
fields since the remaining components give null contribution in
the approach used here \cite{GLENDENNING}.

The $\delta$ and $\xi$ particles can be identified with the $a_0$
(980) and $f_0$(980) states, respectively. The $\xi$ and $\phi$
states are assumed as mainly composed by a $s \bar{s}$ pair and
therefore they couple only to the hyperons.  Furthermore the
values
$m_\phi=1020$ MeV, $m_\xi=975$ MeV are adopted.\\
The Lagrangian (\ref{LAGRANGIAN}) belongs to the framework of the
Quantum Hadrodynamics \cite{WALECKA}. The original model was
progressively completed with additional meson-meson interactions
with the aim to improve the agreement with the phenomenological
knowledge. Thus, for instance, a better result for the
compressibility was obtained by the inclusion of polynomial
self-interaction in the scalar iso-scalar $\sigma$ meson
\cite{BOGUTA}. The density dependence of the symmetry energy was
adjusted through vertices mixing the vector mesons $\omega-\rho$
\cite{HOROW1}, as well as quartic $\omega$ self-interaction was
tuned to reproduce the maximum mass of a neutron star
\cite{HOROW2}. More recently a mixing term of the scalar
$\sigma-\delta$ mesons was proposed to improve the results for the
radius of a neutron star and the tidal deformability of a binary
system \cite{MCS}. It must be noted that this development has
focused on physical systems with increasing energy and matter
densities as they become accessible to the experimental evidence.
This work continues this line of thought in the context of the
hyperon puzzle. Two additional mixing vertices, not contemplated
previously, are proposed here. They involve the scalar meson $\xi$
which constitutes an exclusive channel for the hyperon
interaction, and they reinforce the feedback with the nucleon
fields.

The coupling constants $g_{m b}, m=\sigma, \xi, \delta, \omega,
\rho, \phi$ and $ A, B, G_{\sigma \delta}, G_{\sigma \xi}, G_{\xi
\delta}, G_{\omega \rho}$  are fixed to reproduce a set of
selected empirical data. The $\delta$ and $\rho$ fields, in
contrast with the other mesons, couple differently to each member
of a baryon iso-multiplet since $g_{\delta b},\,g_{\rho b}$ are proportional
to the third projection of the isospin number of the baryon $b$.\\
The equations of motion corresponding to this Lagrangian are
solved in the mean field approximation for uniform dense matter,
in a reference frame where the mean value of the spatial component
of the baryon currents are zero. Furthermore, all the degrees of
freedom are considered as stable states of the strong interaction.
Under such conditions the equations are greatly simplified, since
the meson mean values do not vary spatially, and only the third
component of the meson iso-multiplets are non-zero
\begin{equation}
\left(i \not \! \partial -M^*_b - g_{\omega b} \, \omega_0 -
g_{\phi b}\,  \phi_0 - g_{\rho b}\,  \rho_0 \right)\psi_b=0,
\nonumber \end{equation}
\begin{eqnarray}\left( m_\sigma^2-2\, G_{\sigma \delta}\,\delta^2-2\, G_{\sigma \xi}\,\xi^2 \right) \sigma +
A \sigma^2 + B \sigma^3&=& \sum_b g_{\sigma b}\, n_{sb},\nonumber \\
\left( m_\delta^2-2\, G_{\sigma \delta}\,\sigma^2 -2\, G_{\xi \delta}\,\xi^2\right) \delta &=& \sum_b g_{\delta b}\, n_{sb},\nonumber \\
\left(m_\xi^2-2\, G_{\sigma \xi}\,\sigma^2-2\, G_{\xi
\delta}\,\delta^2\right) \xi &=& \sum_b g_{\xi b}\,
n_{sb},\nonumber
\end{eqnarray}
\begin{eqnarray}\left(m_\omega^2
+ 2 \,G_{\omega \rho}\ \rho_0^2\right)
\omega_0
& =& \sum_b g_{\omega b} \,n_b \nonumber\\
\left( m_\rho^2+ 2\, G_{\omega \rho}\, \omega_0^2\right) \rho_0
&=& \sum_b g_{\rho b}  \,n_b, \nonumber \\ m_\phi^2 \,\phi_0&=&
\sum_b g_{\phi b} n_b \nonumber
\end{eqnarray}
where $M^*_b=M_b-g_{\sigma b}\, \sigma- g_{\xi b} \, \xi-
g_{\delta b}\,  \delta$ is the effective mass of the baryon $b$,
and the source of the meson equations are the baryon densities
\[
n_b=  \frac{p_b^3}{3 \pi^2} ,\; \; \; \;n_{sb}= \frac{M^\ast_b}{2
\pi^2} \left[ p_b E_b- M^{\ast 2}_b \ln \left(\frac{p_b+
E_b}{M^\ast_b} \right)\right]\nonumber
 \]
 The left side equation introduces the Fermi momentum $p_b$, and
 $E_b=\sqrt{p_b^2+M^{\ast 2}_b}$ is used. Within the approach, the energy density of the system is given by
\begin{eqnarray}
\mathcal{E}_H&=&\frac{1}{4} \sum_b \left(n_{s b} M^\ast_b + 3 n_b
E_b \right)+\frac{1}{2}\left( m_\sigma^2 \sigma^2 + m_\delta^2
\delta^2 +m_\xi^2 \xi^2 +m_\omega^2 \omega_0^2+ m_r^2 \rho_0^2+
m_\phi^2 \phi^2 \right) \nonumber\\
&&+ \frac{A}{3}
\sigma^3+\frac{B}{4} \sigma^4 - G_{\sigma \delta} \sigma^2
\delta^2- G_{\sigma \xi} \sigma^2
\xi^2- G_{\xi \delta} \xi^2
\delta^2+ 3 G_{\omega \rho} \,\omega_0^2\,\rho_0^2  \nonumber
\end{eqnarray}
The pressure is obtained by the canonical relation at zero
temperature \[P=\sum_b \mu_b\, n_b - \mathcal{E}_H,\] and the
chemical potentials are given by $\mu_b=E_b+ g_{\omega
b}\,\omega_0 + g_{\phi b}\,\phi_0+ g_{\rho b}\rho_0$.

The bi-quadratic coupling between scalar mesons has been
considered  in \cite{KUBIS} with the aim of studying how the
properties of the neutron stars are affected by the mixing of the
$\sigma - \delta$ scalar mesons.  The model was studied in
\cite{MCS}, where the reference state for nuclear matter was fixed
at the baryonic density  $n_0=0.16$ fm$^{-3}$ and zero
temperature. At such point the following empirical values
$E_\text{bind}=-16$  MeV, $E_\text{sym}=32$ MeV, $M_N^*/M_N=0.65$,
$K=230$ MeV,  and $L=50$ MeV were adopted for the binding energy,
the symmetry energy, the effective nucleon mass ($N=n,\, p$), the
nuclear compressibility and the slope parameter of the symmetry
energy respectively.  They serve as constraints to determine the
constants $g_{\sigma N}, g_{\omega N},  A$ and $B$, and to give a
reasonable range of variation for $g_{\delta p}, g_{\rho p},
G_{\omega \rho}$ and $G_{\sigma \delta}$. An examination of the
predictions for the tidal deformability $\Lambda_{1.4}$ has shown
that larger values of $G_{\sigma \delta}$ improves the agreement
with the constraints provided by the GW170817  event \cite{MCS}.\\
A further extension was presented in \cite{AGUIRRE} introducing
hyperons as well as the $\phi$ vector meson. The  couplings
between hyperons and vector mesons were fixed according to the
SU(6) symmetry of the quark model
\[ g_{\omega \Lambda}=g_{\omega \Sigma}=2\,g_{\omega \Xi}=\frac{2}{3} g_{\omega
N},\]
\[ g_{\rho \Lambda}=0, \;\frac{1}{2}\, g_{\rho \Sigma^+}=g_{\rho \Xi^0}=-\frac{1}{2}\,g_{\rho \Sigma^-}=-g_{\rho \Xi^-}=
\,-g_{\rho n}=\,g_{\rho p},\]
\[g_{\phi \Lambda}=g_{\phi \Sigma}=\frac{1}{2}\,g_{\phi \Sigma}=-\frac{\sqrt{2}}{3}\,g_{\omega N}.\]

The three parameters $g_{\sigma b},\, b=\Lambda, \Sigma, \Xi$ are
determined by adjusting the energy $U_b=g_{\omega b}\,
\omega-g_{\sigma b}\, \sigma$  of an isolated hyperon at rest,
immersed in isospin symmetric nuclear matter at the normal
density. Although their empirical values are poorly known, they
are usually taken as
\begin{equation} U_\Lambda=-30\, \text{MeV}, \;U_\Sigma=30\, \text{MeV}, \,
U_\Xi=-18 \, \text{MeV}. \label{STANDARD}\end{equation}
The values $g_{\sigma \Lambda}=5.616$, $g_{\sigma \Sigma}=3.989$,
and $g_{\sigma \Xi}=2.920$ are thus obtained.
In contrast to the simplifications made in \cite{AGUIRRE}, the
coupling between the hyperons and the scalar mesons $\xi$ and
$\delta$ are fully considered here. The latter distinguishes the
components of a baryonic isospin multiplet, and particularly does
not couple to those hyperons with zero third isospin component
$I_b=0$, hence $g_{\delta \Lambda}=g_{\delta \Sigma^0}=0$. In
addition the assumption that the strength of its coupling to a
given baryon decreases with its strangeness content leads one to
$2\,g_{\delta p}/3=g_{\delta \Sigma^+}=g_{\delta \Xi^0}$, opposite
signs must be assigned to the  complementary isospin projections
$g_{\delta n}=-g_{\delta p}$, $g_{\delta \Sigma^-}=g_{\delta
\Xi^-}=-2\,g_{\delta p}$. The vertices involving the $\xi$ meson
are exclusive for the hyperons, hence they could be deduced from
the scarce information on hyperon matter \cite{SCHAFFNER,VIDANA}.
As a phenomenological guide one can consider the criterium that
relates the excess binding energy $\Delta B(A)$ of a double
$\Lambda$ hypernucleus of atomic number $A$ to the single particle
potential $U_\Lambda^{(\Lambda)}(n)$ in $\Lambda$ matter at
density $n$, i. e.
\[\Delta B(A)\simeq U_\Lambda^{(\Lambda)}(n)-2\, U_\Lambda^{(\Lambda)}(2\,n)\simeq -U_\Lambda^{(\Lambda)}(n)\]
where $n \simeq n_0/A$ \cite{VIDANA}, and $\Delta B(6)=0.67$ MeV
is identified as the experimental value of the $_{\Lambda \Lambda}
^6He$ hypernucleus, see for instance \cite{GAL}. Thus the
constraint $U_\Lambda^{(\Lambda)}(n_0/5)=-0.67$ MeV is adopted
here in agreement with other investigations on compact stars
\cite{OERTEL,TORRES, LILONG}.\\
Following the arguments of \cite{SCHAFFNER}, the single particle
potentials of the $\Xi$ and $\Lambda$ hyperons are related by
\[U_\Xi^{(\Xi)}(n_0)=2 \, U_\Lambda^{(\Lambda)}(n_0/2)\]
The strategy to fix the remaining constants of the model is to
explore the bidimensional space $G_{\sigma \xi}\;G_{\xi \delta}$
with the aim of accommodating neutron stars with mass $M \simeq 2
M_\odot$. In the process it was found that the value of $G_{\xi
\delta}$ is irrelevant due to the smallness of the product $\xi^2
\,\delta^2$, therefore $G_{\xi \delta}=0$ is taken from here on.
The numerical values of the remaining parameters are as follows:
$m_\sigma=500$ MeV, $m_\delta=983$ MeV, $m_\omega=783$ MeV,
$m_\rho=770$ MeV, A=13.08 fm$^{-1}$, B=-31.6, $g_{\sigma N}=9.22$,
$g_{\omega N}=11.35$, $g_{\delta p}=\sqrt{5.2\, \pi}$, $g_{\rho
p}= \,\sqrt{11.08 \pi}$, $G_{\sigma \delta}=50$, $G_{\sigma
\xi}=-100$, $G_{\omega \rho}=263.92$, $g_{\xi \Lambda}=2.117$, and
$g_{\xi \Xi}=10.077$. The criterium to select this specific value
for $G_{\sigma \xi}$ is discussed in Sec.\ref{Sec4}.

The hypothesis of homogeneous matter which leads to the equations
of motion shown above, is appropriate for densities greater than
several tenths of the normal nuclear value $n_0$. The
electromagnetic interaction, not included in (\ref{LAGRANGIAN}),
gives rise to non-homogeneous structures. For this reason the EoS
evaluated in \cite{BPS} is adopted for the low density regime and
assembled to the results of the interaction
(\ref{LAGRANGIAN}) by imposing continuity at the matching point $n=0.35 n_0$.\\
For very dense matter it is expected that hadrons are not longer
the most stable configuration and a transition to deconfined
quarks happens. To take account of this state of homogeneous quark
matter, the NJL model is implemented with inclusion of a vector
interaction \cite{KLEVANSKY}. The NJL presents interacting quarks
which generate their own constituent masses. This effective mass
depends on the properties of the medium and are expected to
decrease with increasing baryonic density. \\
The energy density is given by
\begin{equation}
\mathcal{E}_\text{Q}=\sum_q \left[\frac{N_c}{\pi^2}
\int_\Lambda^{p_q} dp\,p^2 \, \sqrt{p^2+M_q^2}+ 2 \, G \, n_{s q}+
18 \, G_v \, n_q^2\right]- 4\, K\, n_{s u} \, n_{s d }\, n_{s s}+
\mathcal{E}_0 \label{Enjl}
\end{equation}
where $q=u,\,d,\, s$, $p_q$ is the Fermi momentum which is related
to the baryonic number density by $n_q =p_q^3/3\pi^2$. \\
The effective masses are given by
\[M_i=m_i-4\, G\, n_{s i}+2\, K \, n_{s j}\, n_{s k},\; j\neq i \neq k.  \]
where $m_q$ is the current quark mass, and the quark condensates
$n_{s q}$ can be expressed as
\[ n_{s q}=\frac{N_c}{\pi^2} \,M_q \int_\Lambda^{p_q} \frac{dp \, p^2}{\sqrt{p^2+M_q^2}}\]

A cutoff $\Lambda$ is used to renormalize ultraviolet divergences
in the momentum integration, $\mathcal{E}_0$ is a constant
introduced to obtain zero vacuum energy, $G$ and $K$ are the
couplings for four and six quark interactions, and $G_v$ is the
strength of the vector current-current interaction. The chemical
potential for each flavor is simply $\mu_q=\sqrt{p_q^2+M_q^2}+12
\, G_v n_q$.\\
For the numerical calculations the set of constants specified in
\cite{REHBERG} are used. The vector coupling $G_v$ has not been
determined with precision and it is usually taken as a parameter
within the range $0 \leq G_v \leq G$. Therefore the relatively low
value $G_v=0.08\, G$ is chosen here in order to obtain a threshold
density $n \geq 4\, n_0$ for the deconfinement process.

The transition between the hadronic and deconfined phases has been
described within different dynamical schemes. In this work the
picture of a continuous and monotonous EoS, with an intermediate
state of coexisting phases is adopted. It is commonly denominated
as the Gibbs construction. If $\chi$ is the spatial fraction
occupied by the deconfined phase, then the total energy and the
baryonic number densities of the system are given by
\begin{eqnarray} \mathcal{E}&=&\chi\,\mathcal{E}_\text{Q} + (1-\chi)\, \mathcal{E}_\text{H},
\\
n&=&\chi\, n_\text{Q} + (1-\chi) \,  n_\text{H}
\label{BDens}\end{eqnarray}

Furthermore, for thermodynamical equilibrium of the coexisting
phases the partial pressures of each phase must coincide
\begin{eqnarray} P_B=\sum_b \mu_b\,n_b - \mathcal{E}_\text{H}=
\sum_q \mu_q\,n_q - \mathcal{E}_\text{Q}\label{PCond}.
\end{eqnarray}

To describe neutron star matter the complementary requirement of
electrical neutrality is imposed. To reach this condition a fluid
of non-interacting leptons (electrons and muons) is considered,
which freely distributes among the hadron and quark phases so that
the condition
\begin{equation}0=3\,\chi \sum_q C_q\,n_q + (1-\chi)
\sum_b C_b \,n_b -\sum_l n_l,\label{ECharge}\end{equation}
is satisfied. In this expression $C_k$ stands for the electric
charge in units of the positron charge.\\
These leptons also contribute to the total energy by
\[\mathcal{E}_\text{L}=\frac{1}{\pi^2}\sum_l \int_0^{p_l} dp\,p^2
\, \sqrt{p^2+m_l^2} \] where $n_l=p_l^3/3 \pi^2$, their chemical
potentials can be written as $\mu_l=\sqrt{p_l^2+m_l^2}$, and the
partial lepton contribution to the pressure is $P_L=\sum_l
\mu_l\,n_l - \mathcal{E}_\text{L}$. Hence, the complete
expressions for the energy and the pressure in the mixed phase are
\begin{eqnarray} \mathcal{E}&=&\chi\,\mathcal{E}_\text{Q} + (1-\chi)\, \mathcal{E}_\text{H}+\mathcal{E}_\text{L},
\label{TotalE}\\
P&=&\mu_\text{B}\, n-\mathcal{E}=P_\text{B}+P_\text{L}
 \label{PRESSURE}.\end{eqnarray}

 The coefficient $\chi$ is obtained by using the conditions of conservation of
 the baryonic number, the electric charge, and thermodynamical
 equilibrium, Eqs. (\ref{BDens}), (\ref{ECharge}) and (\ref{PCond}) respectively. Thus, it is uniquely
 determined for each density of neutral matter in equilibrium at zero
 temperature, and it is a dynamical property of the combination of models
 used.

There are two conserved charges which characterize the global
state of the system, the baryonic number and the electric charge
with associated chemical potentials $\mu_\text{B}$ and
$\mu_\text{C}$ respectively. It must be noted that the last one
does not enter in the intermediate expression of Eq.
(\ref{PRESSURE}) because the total electric charge is zero. Both
chemical potentials can be combined to give the chemical
potentials of all baryons, quarks and leptons circumstantially
present. Therefore they are linearly dependent through the
relations of equilibrium against beta decay.

\section{Properties of the neutron star}\label{Sec3}

The structure of an isolated neutron star can be solved using the
Tolman-Oppenheimer-Volkov equations for the spherically symmetric
case
\begin{eqnarray}
\frac{dP}{dr}&=&- \, [{\cal E}(r)+P(r)]\,[{\cal M}(r)+4 \pi
r^3 P(r)]\,\frac{e^{2 \lambda(r)}}{r^2} \, , \nonumber \\
{\cal M}(r)&=&\int_0^r 4 \pi \, {r'}^2 \, {\cal E}(r') \, dr' \, .
\nonumber
\end{eqnarray}
Units for which $c=1, \,G=1,\, \hbar=1$ has been used. The
relation ${\cal E}(P)$ is provided by the EoS described in the
previous section and the definition
\[e^{\lambda(r)}=\left[1-2  {\cal M}(r)/r\right]^{-1/2}\]
is used.\\
 Starting from given values of
the central pressure and energy, these equations are integrated
outward until a radius $R$ is reached for which  $P(R)=0$, and the
total mass is defined as
$M={\cal M}(R)$. \\
Once the mass ${\cal M} (r)$ and pressure $P(r)$ distributions
inside the star have been determined one can evaluate its moment
of inertia $I$ assuming slow and homogeneous rotation with angular
velocity $\Omega$. By solving the differential equation for
$\varphi(r)=1-\omega(r)/\Omega$, where $\omega(r)$ is the angular
velocity distribution of a fluid element inside the star
\cite{HARTLE}
\begin{equation}\frac{d}{dr}\left[r^4\,j(r)\,\varphi'(r)\right]+4\,r^3\,j'(r)\,\varphi(r)=0,
\; r<R \label{Phi}\end{equation}
with the definition $j(r)=\exp-\left[\lambda(r)+\nu(r)\right]$,
and the metric function $\nu(r)$ satisfies the equation
\[\frac{d\nu}{dr}=-\frac{dP/dr}{{\cal E}(r)+P(r)}\]
and the auxiliary condition $e^{\nu(R)}=e^{-\lambda(R)}$. \\
The solution outside the star is
\[\varphi(r)=1-\frac{2\,I}{r^3},\;r>R.\]
Thus, Eq. (\ref{Phi}) is complemented with the boundary conditions
$\varphi'(0)=0, \varphi(R)=1-2\,I/R^3$.

The tidal deformability of a compact star can be written in terms
of the second Love number $k_2$ as $\Lambda=2\,k_2/3 x^5$, where
the compactness parameter is $x=M/R$. To evaluate the Love number
the radial function $y(r)$, related to the tidal field, must be
found by solving the differential equation
\[ y'(r)+y^2(r)+f(r)\,y(r)+q(r)\,r^2=0\]
subject to the condition $y(0)=2$. The following definitions has
been used
\begin{eqnarray}
f(r)&=&\left[1+4\,\pi\,r^2\left(P-{\cal E}\right)\right]\,e^{2 \lambda(r)} \nonumber\\
q(r)&=&\left[4 \pi \left(5\,{\cal E}+9\,P+\frac{P+{\cal
E}}{v_e^2}\right)-\frac{6}{r^2}-\frac{4}{r^4}\frac{\left({\cal
M}+4\,\pi\,r^3 P\right)^2}{1-2\,{\cal M}/r}\right]\,e^{2
\lambda(r)} \nonumber
\end{eqnarray}
The relativistic speed of sound $v_e$ has been introduced, which
is defined by
 \begin{equation}v_e^2=c^2\,\frac{dP}{d{\cal E}}.\label{EqSpeed}\end{equation}
Finally, the Love number is given by
\begin{eqnarray}k_2&=&\frac{8}{5} x^5 (1-2 x)^2 [2-y_R+2 x (y_r-1)]/\Big\{6 x [2-y_R+x (5 y_R-8)]
\nonumber \\&+& 4 x^3 [ 13-11 y_R+x (3 y_R-2)\nonumber \\&+&2 x^2
(1+y_R)]+3 (1-2 x)^2[2-y_R+2 x (y_R-1)]\ln(1-2 x)\Big\}\nonumber
\end{eqnarray} where $y_R=y(R)$.

In order to simplify the numerical evaluation of the oscillatory
motion inside the star an approximation, known as the relativistic
Cowling approach will be used. It consists in neglecting the
variation of the metric functions of the space-time supporting the
vibrations \cite{McDERMOTT}. The discrepancy between full
calculations and those using the Cowling treatment was studied
long time ago \cite{YOSHIDA1}, and repeated under various
conditions \cite{YOSHIDA2,SOTANItk}. These investigations found
that the difference is of the order $10 - 30 \,\%$. This work
intends a qualitative description for stars with masses greater
than $2 M_\odot$, therefore this approach is adequate and a
discrepancy of about $20 \%$ in the oscillation
frequencies is expected.\\

Assuming spherical symmetry for the equilibrium state, the
displacement of a fluid element located at point $r, \theta, \phi$
 at time $t$ can be decomposed in a multipolar expansion as
\begin{equation}\delta
r=\frac{e^{-\lambda}}{r^2}\,W(r)\,Y_l^m(\theta,\phi)\,e^{i \omega
t},\label{TRANSV}\end{equation}
\begin{equation}\delta \theta=-\frac{V(r)}{r^2}\,\frac{\partial}{\partial
\theta}Y_l^m(\theta,\phi)\,e^{i \omega t},\end{equation}
\begin{equation}\delta
\phi=-\frac{V(r)}{r^2\sin^2 \theta}\,\frac{\partial}{\partial
\phi}Y_l^m(\theta,\phi)\,e^{i \omega
t}.\label{RADIAL}\end{equation}
In this approach the unknown functions $W(r), V(r)$ satisfy the
differential equations
\[\frac{d V}{d r}-2 \frac{d \nu}{d r}\,V+\frac{e^\lambda}{r^2}\,W-\left(\frac{1}{v_e^2}
-\frac{1}{v_a^2}\right)\left(V+e^{2 \nu-\lambda}\frac{d \nu}{d
r}\,\frac{W}{(r \omega)^2}\right)\,\frac{d \nu}{d r}=0,\]
\[\frac{d W}{d r}+l(l+1)\,e^\lambda V-\frac{1}{v_a^2}\left(r^2 \omega^2\,e^{\lambda-2 \nu}
V+ \frac{d \nu}{d r}\,W\right)=0\]
and they are subject to the conditions
\[W(r)+l\, r\,V(r)\rightarrow 0, \text{as}\; r\rightarrow 0,\]
\[R^2\omega^2\, V(R)+\nu'(R)\,e^{3\nu(R)} W(R)=0.\]
The equations above contain the adiabatic speed of sound, defined
by
\begin{equation} v_a^2=c^2\,\frac{\partial P}{\partial {\cal
E}},\label{AdSpeed} \end{equation}
 where the partial
derivative is evaluated by keeping constant the relative
population of each fermionic species,  and the constraint of
$\beta$ equilibrium is imposed after evaluation. The velocity is
related to the adiabatic index by the relation $\gamma
P=\left(P+{\cal E}\right)\,v_a^2$. The differences between
$v_e$ and $v_a$ has been previously discussed, as for instance in \cite{JAIKUMAR}.\\
 In our approach the relation between $P$ and $\mathcal{E}$
is monotonous and continuous, however its first derivative, i. e.
the speed of sound $v_e$, presents finite discontinuities at the
threshold of the phase transition.\\

Explicit expressions for $v_e$ and $v_a$ can be found in \cite{AGUIRRE}.\\
The quantity
\[A_+=-\frac{d\nu}{dr}\left(v_e^{-2}-v_a^{-2}\right)\,e^{-\lambda}\]
discriminates convective instability by the condition $A_+>0$.
Furthermore, it is related to the Brunt-V\"{a}is\"{a}l\"{a}
frequency $N$ by
\begin{equation}N^2=-e^\nu\,A_+\,d\nu/dr.\label{BRUNT}\end{equation}
\begin{figure}[t]\vspace{-2cm}\begin{center}
\includegraphics[width=0.7\textwidth]{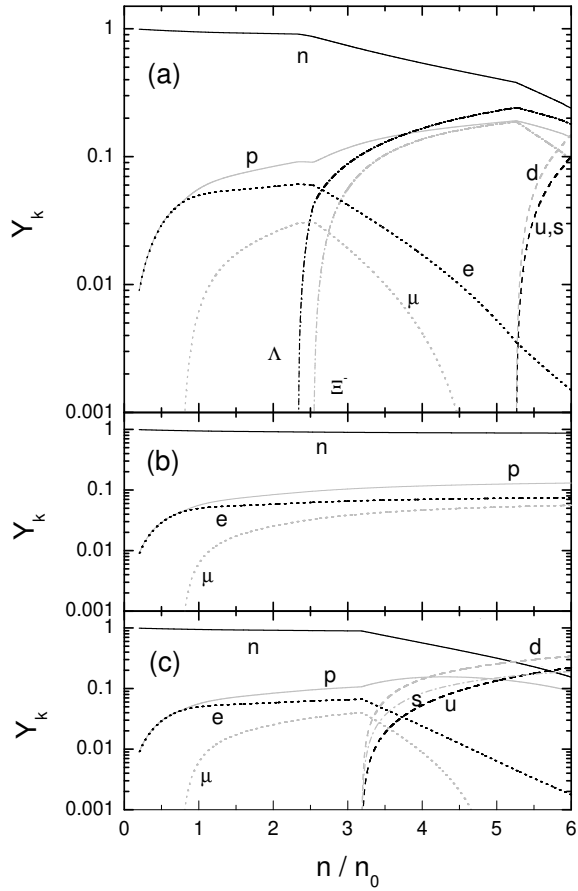}\end{center}
\caption{\footnotesize The partial fraction $Y_k$ for the
different species considered in the approaches F (a), N (b), and
NQ (c). The definitions $Y_b=(1-\chi)\,n_b/n$ for baryons,
$Y_q=\chi\,N_c\, n_q/n$ for quarks, and $Y_l=n_l/n$ for leptons,
are used. The abbreviation F  stand for the case with the full
octet of baryons and quarks, N for the case with only nucleons,
and NQ for the case of nucleons combined with quarks, in agreement
with the convention stated in the main text.}
\end{figure}

\section{Results and discussion}\label{Sec4}

In this section an analysis of the EoS of the model proposed and
its ability to adjust the relevant information provided by the
observational evidence about compact stars is made. It must be
mentioned that the model tries to cover a wide range of
phenomenological results but keeping simple.  The classical
description given in \cite{BPS} takes account of the emergence of
atomic nuclei at low densities.  The covariant field theory of
hadrons is appropriate for medium densities where relativistic
effects becomes important and the homogeneity of matter is a
plausible assumption. Furthermore new hadronic and leptonic
degrees of freedom, relevant for the composition of a neutron
star, are easily included. Finally for the highest densities
achievable in the core of a compact star, it is expected that
hadrons are replaced by deconfined quarks. The NJL framework is
adopted for this stage, since it effectively represents the strong
interaction for the energy regime involved.\\
The determination of the parameters of the model follows this
staggered scheme, the masses and couplings of the conventional
degrees of nuclear physics are fixed using the well established
phenomenology at the normal density $n_0$. More uncertain are the
parameters of the hyperonic interaction due to the lack of precise
experimental data. For instance, the hyperon-nucleon interaction
mediated by the scalar $\sigma$ meson is normalized by the single
hyperon potentials in nuclear matter. The coupling of the hidden
strangeness $\xi$ meson, responsible of the hyperon-hyperon
interaction, is deduced using single hyperon potentials in
hyperonic matter. Other constants are chosen to improve the
agreement with astrophysical data. This is the case of the
constant $G_{\sigma \delta}$, whose preferable value for adjusting
the tidal deformability $\Lambda_{1.4}$ was discussed in
\cite{MCS}. A similar case is found for the vector current-current
coupling of the NJL. Since an increase of $G_v$ has the effect of
enlarging the deconfinement density as well as the central density
while decreasing the radius of the maximum mass star, the value
$G_v=0.08\, G$ was chosen in order to optimize the coincidence
with the result $M_\text{max}\simeq 2\,M_\odot$.\\
In regard of the coupling $G_{\sigma \xi}$ one can expect
numerical values comparable to $G_{\sigma \delta}$ obtained in
\cite{MCS}, with the necessary modification to take account of the
expected smaller amplitude of $\xi$ as compared to $\delta$. Thus
the range $|G_{\sigma \xi}| \leq 100$ has been initially explored.
By changing its numerical value, but keeping the constraint on the
single particle potentials of the hyperons, an increasing trend on
the maximum mass of a neutron star was obtained as $G_{\sigma
\xi}$ is decreased. Thus for $G_{\sigma \xi}=0,\,-20,\,-60$ the
results $M_\text{max}/M_\odot=1.94,\; 1.96,\;1.98$ are obtained
for a particular value of $G_v$.
\begin{figure}[t]\vspace{-2cm}\begin{center}
\includegraphics[width=0.7\textwidth]{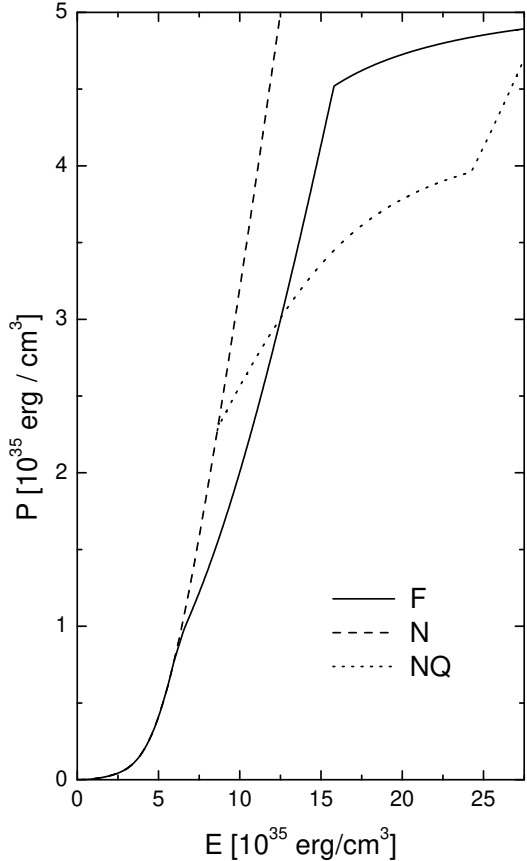}\end{center}
\caption{\footnotesize The equation of state for the composite
model, the different approaches are distinguished according to the
line convention shown. As expected the EoS becomes softer around
the hyperon onset and the deconfinement threshold. Hence the case
N exhibits the steepest EoS, while the curve F keeps harder than
the NQ one because of its delayed deconfinement transition. The
abbreviations F, N, and NQ are described in the caption of Fig.
1.}
\end{figure}

As a first probe the partial fraction of baryons and leptons is
examined in Fig. 1. The results for the full treatment (F) are
displayed in the upper panel, while the bottom one combines the
case with nucleons and leptons only (N), together with that
considering nucleons, leptons, and quarks (NQ). As it is usual,
the onset of the hyperons $\Lambda$ and $\Xi^-$ at densities near
$2.3\,n_0$ causes a pronounced decrease of the population of
leptons, and particularly of the muons that are extinguished
around $n/n_0=4.5$. The fraction of neutrons is also diminished in
contrast with the slight rise shown by the protons. At a higher
density $n/n_0=5.3$ starts the coexistence with a phase of
deconfined quarks, which produce the final decline of all the
baryonic species.  It must be mentioned that within this approach
the core of the more massive neutron star has a considerable
fraction of hadrons, i.e. no pure quark matter is found there.\\
The case N shows that if hyperons and quarks were suppressed, the
composition of beta equilibrated matter becomes asymptotically
stable with increasing density (Fig.1b). However if a continuous
transition from nuclear to quark matter is allowed, as in the NQ
case Fig.1c, the deconfinement happens at a lower density $n/n_0
\simeq 3.2$.

 The EoS obtained in this model is shown in Fig.2.  A low energy regime,
consisting of pure nuclear matter and leptons, is distinguished by
the coincidence of the three cases F, N, and NQ. The $\Lambda$
hyperon becomes stable at a density $n = 2.3\, n_0$, corresponding
to the split of the F and N curves. The $\Xi^-$ also appears
before the coexistence with deconfined quarks takes place. This
fact is marked by a sudden change in the tangential direction of
the curve F. \\
On the other hand, pure nuclear matter becomes unstable earlier,
at the point where the N and NQ curves separate. While the upper
part of the NQ curve shows the end of the coexistence regime. As
expected, the inclusion of configurations that minimize the energy
of the system leads to a softening of the EoS.\\
The pressure at the density $n/n_0=2$ has been estimated in
\cite{ABBOT2} as $P=3.5^{+2.7}_{-1.7} \times 10^{34}$ dyn/cm$^2$
in order to be consistent with the observational data of the
GW170817 event. In the present calculations the result $P=4.2
\times \,10^{34}$ dyn/cm$^2$ has been obtained, which is comprised
within the confidence band.

 An interesting fact can be appreciated in Fig. 3,
which shows the neutron and proton masses as functions of the
density. The presence of the hyperons makes the decrease of the
nucleon masses more pronounced. In particular the neutron mass is
almost collapsed when the deconfinement starts. In the mixed
phase, instead, it increases slightly so that the mass difference
of the duplet is considerably reduced. The same feature is
repeated in the case of only nucleons coexisting with unbound quarks.\\
The behavior of the masses is a consequence of the variance of the
amplitude of the scalar fields, as shown in Fig. 4. The increase
of the $\sigma$ meson has an additional source in the $\sigma^2
\xi^2$ vertex which becomes active at the onset of the $\Lambda$
hyperon (F case). However the $\xi$ mean field decreases in the
mixed phase as the hyperon population diminishes. This effect is
reinforced by the simultaneous decrease of the amplitude of the
$\delta$ meson.\\
In a certain sense the exclusive hyperon-hyperon interaction gives
physical continuity to the model. The exchange of the $\phi$ and
$\xi$ mesons represent the vector
\begin{figure}[t]\vspace{-2cm}\begin{center}
\includegraphics[width=0.8\textwidth]{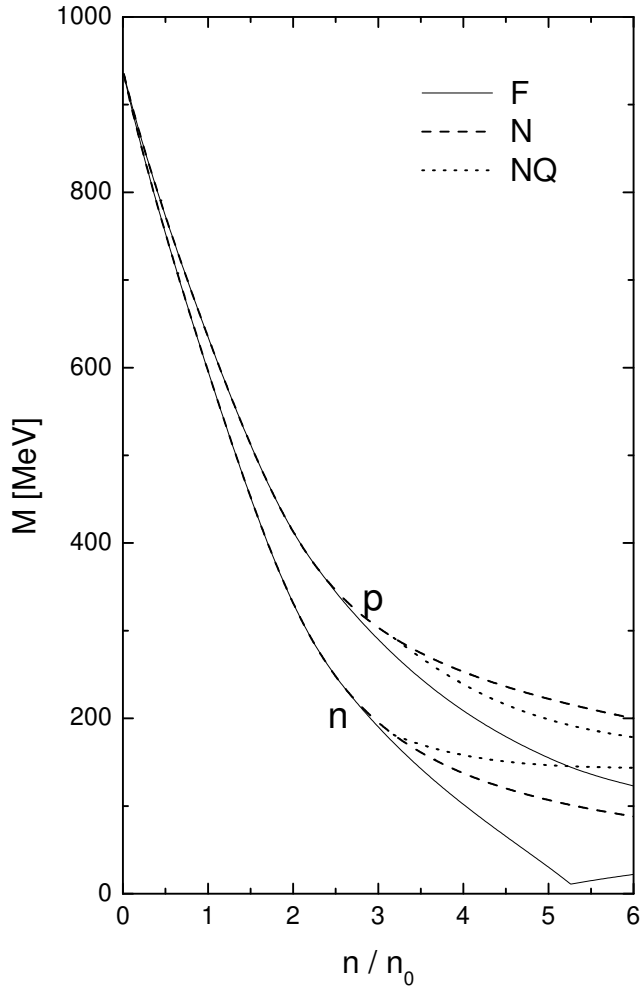}\end{center}
\caption{\footnotesize The mass of the proton-neutron duplet as
functions of the baryonic density for different approaches. For
each of these nucleons the three curves coincide below the hyperon
onset. Above the deconfinement threshold the case F shows the
lightest masses. While for the N and NQ instances, the results are
opposite for protons and neutrons. This is a consequence of the
weakness of the $\delta$ meson in the coexistence phase, favoring
the degeneracy of the masses of the nucleons. The abbreviations F,
N, and NQ are described in the caption of Fig. 1.}
\end{figure}
and scalar channels contributing
to the repulsive and attractive terms of the EoS, respectively.

These contributions are balanced by reproducing the empirical data
proposed for multi-hyperon systems \cite{SCHAFFNER,VIDANA,GAL}.
The former promotes the deconfinement transition by matching
opportunely the EoS provided by the NJL. The latter favors the
high density regime of the nucleon masses shown in Fig. 3,
mimicking the dynamical recovery of chiral symmetry. Although this
is not the real case because the coupling between nucleons and the
scalar mesons are still active. Hence if one tries $G_{\sigma
\xi}=0,\,g_{\phi Y}=0,\,g_{\xi Y}=0$ then the deconfinement
transition will not be reached, and the trends of the hadronic
results will continue until the collapse of the neutron mass
determines the limit of applicability of the model. One can
conclude that in the present work the two models approach is
necessary and they are really complementary.
\begin{figure}[t]\vspace{-2cm}\begin{center}
\includegraphics[width=0.7\textwidth]{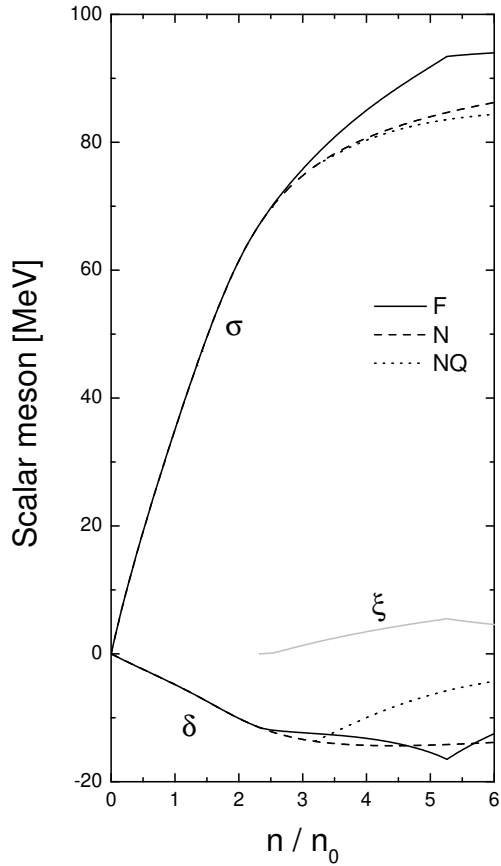}\end{center}
\caption{\footnotesize The amplitude of the mean field solutions
for the scalar mesons $\sigma, \delta$, and $\xi$ as functions of
the baryonic density for different approaches. It is remarkable
the strengthening of the $\sigma$ meson (case F) beyond the
hyperon onset due to the feedback with the $\xi$. While the
coexistence of phases causes a decrease of the amplitude of all
the mesons (cases F and NQ).}
\end{figure}
As discussed in the previous paragraphs, the parameters of the
model can not be completely determined by the experimental
constraints and there is some extent of freedom in their choice.
Some judicious assumptions have been made in the present work, and
the effect of their variance deserves further investigation.

The EoS just described are used as input to solve the macroscopic
properties of a non rotating neutron star, as for instance the
mass-radius relation shown in Fig. 5. The maximum masses are
$M/M_\odot=2.01,\, 2.45,$ and $2.17$, whereas the corresponding
radius are $R=11.7,\, 11.7,$ and $12.5$ in the F, N, and NQ
treatments, respectively. The central density of such stars are
$n/n_0=6.35, \, 5.33,$ and $4.95$, so that the core of the star
does not support pure quark matter. However, in the F and NQ cases
it is composed of an admixture of hadrons and deconfined quarks.\\
The radius of a neutron star could bring important information on
the underlying EoS \cite{ZHANGLI}, therefore it is interesting to
make a comparison with  the Bayesian analysis presented in
\cite{RILEY} for the massive pulsar PSR J0740+6620. It obtains the
radius $R=12.39^{+1.30}_{-0.98}$ for the star with
$M/M_\odot=2.072^{+0.067}_{-0.066}$. The main result of this work,
i.e. F case, is compatible with these constraints and therefore
the presence of hyperons as well a deconfinement transition are
not ruled out by them.\\
\begin{figure}[t]\vspace{-2cm}\begin{center}
\includegraphics[width=0.7\textwidth]{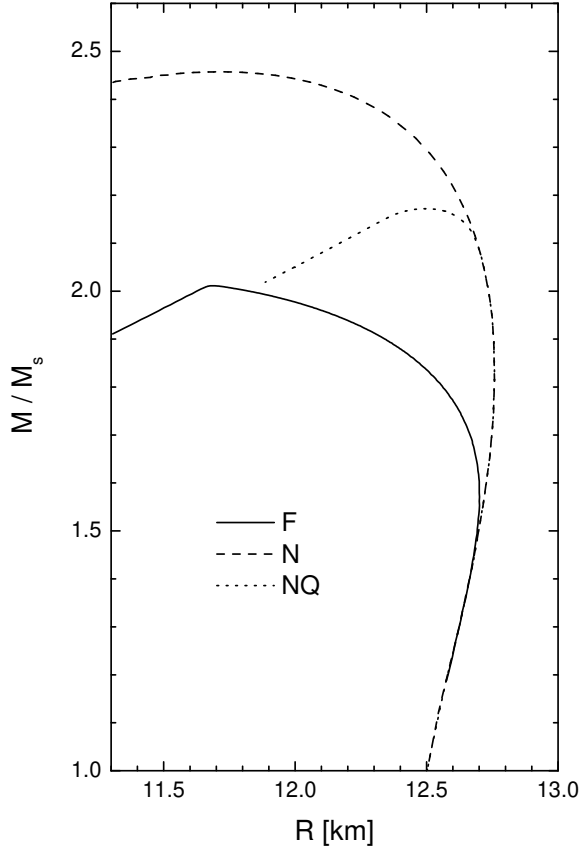}\end{center}
\caption{\footnotesize The mass-radius relation for neutron stars
for different approaches. As expected, the greatest mass
corresponds to the case where only conventional degrees of freedom
are used (N). When a transition to deconfined quarks is allowed,
$M_\text{max}$ decreases but the corresponding radius is
relatively large (NQ). In the full treatment $M_\text{max}$ is
slightly greater than two solar masses (F). The abbreviations F,
N, and NQ are described in the caption of Fig. 1.}
\end{figure}
In regard of the neutron star with the standard $M/M_\odot \simeq
1.4$ mass, in the present work it is found mainly composed by
nucleons and a tiny $1 \%$ of $\Lambda$ hyperons in the core of
the star. The corresponding radius $R_{1.4}=12.66$ km  can be
contrasted with the estimates $12.33^{+0.76}_{-0.81}$ km and
$12.18^{+0.56}_{-0.79}$ km obtained by different approaches in
\cite{RAAIJMAKERS}, or the result $R_{1.4}=12.45 \pm 0.65$ km
obtained in \cite{MILLER}.

At this point a general comparison with previous works using
similar conceptual tools as in the present investigation is
opportune. All of them use a hadronic field model with hyperons
and combined with NJL in a continuous phase transition. The
partial contribution of baryons and quarks to the composition of
stellar matter is in qualitative coincidence with the calculations
shown in \cite{YANGSHEN}. This reference considers two versions of
the NL3 (TM1) parametrization, obtaining  a deconfinement
threshold located around $n_t/n_0=3.1 \;(4.4)$, the maximum mass
is approximately $2.03 \,M_\odot$ ($1.65 \,M_\odot$) and the
corresponding radius $R\simeq 13.6$ km ($R\simeq 13.3$ km). In all
the cases the fiducial star with $M/M_\odot=1.4$ has a large
radius $R_{1.4}>14.2$ km. The same quantities as evaluated in
\cite{HAN} are $n_t/n_0=2.3$, $M_{\text max}/M_\odot \simeq 2.2$
and $R \simeq 13.2$ km, however the authors discard this approach
due to the impossibility to adjust reasonable values for the tidal
deformability. In addition the outcomes of \cite{KUMAR MISHRA}
place the transition at $n_t/n_0=2.7$, besides $M_{\text
max}/M_\odot \simeq 2.55$,  $R \simeq 14.2$ km and $R_{1.4}=14.5$
km.\\
As a partial conclusion it can be stated that the present
treatment produces the highest threshold density for the
deconfinement transition, noticeably improves the result for
$R_{1.4}$ and gives a reasonable maximum mass.\\
\begin{figure}[t]\vspace{-2cm}\begin{center}
\includegraphics[width=0.8\textwidth]{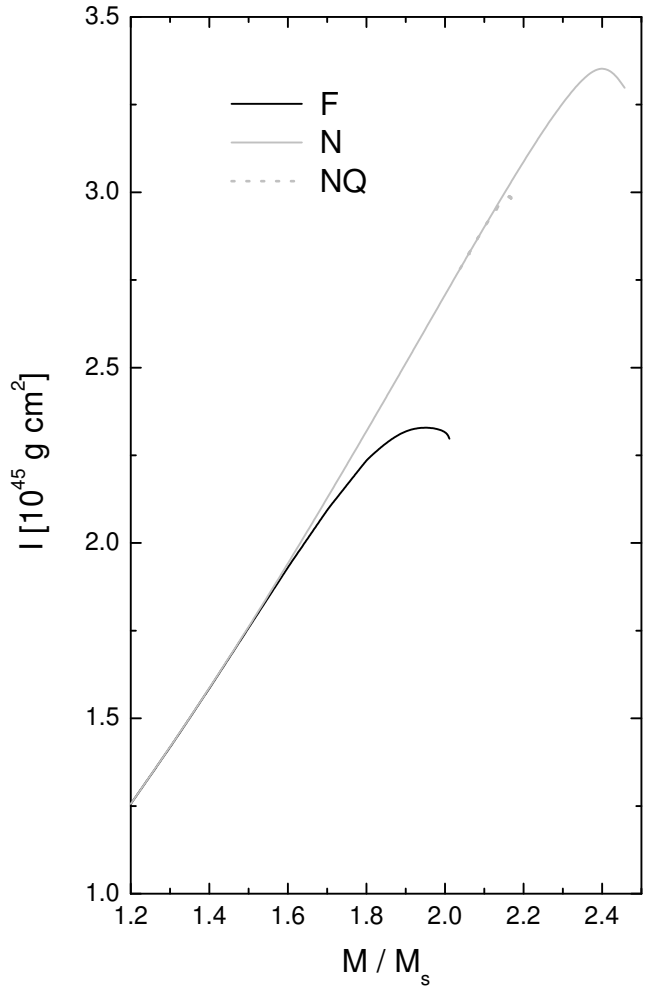}\end{center}
\caption{\footnotesize The moment of inertia of a slowly rotating
neutron star in terms of the star mass  for different approaches.
This quantity is defined as $I=R^3\left[1-\varphi(R)\right]/2$,
where $\varphi$ is the solution of the differential equation
(\ref{Phi}). The curve NQ is almost indistinguishable from the
case N, but it finishes below $n/n_0 \simeq 2.2$.}
\end{figure}
 The effect of the mixing vertex for the $\sigma-\xi$ mesons
is clarified by contrasting with the outcomes of the related model
used in \cite{AGUIRRE}. The deconfinement threshold density is
shifted towards higher values, a significant increase of almost
$26 \%$ is obtained in the F case.  The properties of the maximum
mass star, instead, are slightly modified. The value for $M_{\text
max}$ is raised by less than $1 \%$, whereas the corresponding
radius is decreased by a $6 \%$. The main features of the fiducial
star remain practically unchanged because of the tiny fraction of
hyperons present in such case.

The moment of inertia of a slowly rotating star is presented in
Fig. 6 as a function of the inertial mass. The presence of
hyperons is discernible only for stars with $M\simeq 2\, M_\odot$
due to the fact that the moment of inertia takes a maximum value
before $M_\text{max}$ is reached.  The candidate for an imminent
precision measurement PSR J0737-3039A has a mass
$M/M_\odot=1.338$, for which the analysis made in \cite{LIM}
suggest $I=1.36^{+0.15}_{-0.32}\times \,10^{45}$ g cm$^2$. In the
present calculations the result is $I=1.48\times \,10^{45}$ g
cm$^2$, which is slightly greater than the most probable value
given there, but within the confidence band. The same conclusion
holds respect to the prediction $I=1.15^{+0.38}_{-0.24}\times
\,10^{45}$ g cm$^2$ given in \cite{LANDRY}. It must be mentioned
that within the model used here, such star would have a central
density $n\simeq 2.3 n_0$ and
therefore would be composed only by nucleons and leptons.\\
\begin{figure}[t]\vspace{-2cm}\begin{center}
\includegraphics[width=0.8\textwidth]{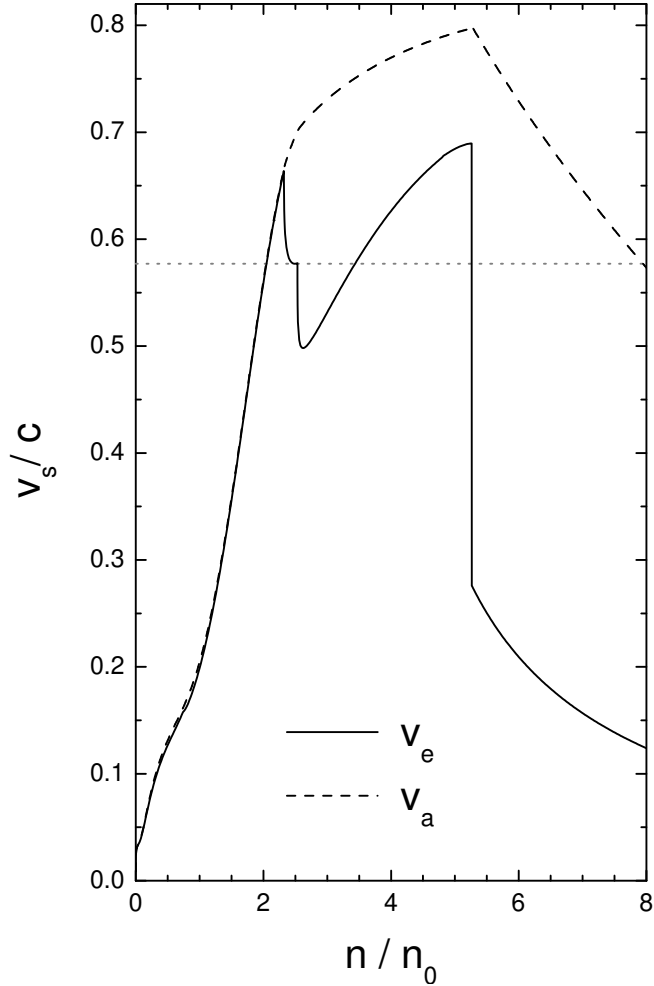}\end{center}
\caption{\footnotesize The equilibrium $v_e$ and adiabatic $v_a$
speeds of sound as functions of the baryonic density in the F
case. The horizontal line corresponds to the conformal limit. The
definitions of both speeds are given in Eqs. (\ref{EqSpeed}) and
(\ref{AdSpeed}).}
\end{figure}
Calculations of the adimensional moment $\tilde{I}=I/M\,R^2$ in
terms of the compactness $x=M/R$ can be contrasted with the
phenomenological relation $\tilde{I}(x)$ found in \cite{LATTIMER}.
Considering the cases of the stars with $M=1.338 M_\odot$, and $2
M_\odot$ discrepancies of only $1\%$ and $3 \%$ have been found
respectively. Similar results are obtained by comparing with the
formula presented in \cite{LIM}.

The rich structure of the speed of sound is closely related to the
symmetry energy of nuclear matter \cite{BAOLI}, the result of the
present model is shown in Fig. 7 in terms of the baryonic density.
As already noted in \cite{JAIKUMAR}, the adiabatic velocity $v_a$
is continuous but $v_e$ presents finite discontinuities at the
beginning of the coexistence domain, see for instance
\cite{PODDER}. Both definitions are almost coincident for low
densities. In fact, if the homogenous matter assumption is
extended for $n \rightarrow 0$, then pure neutron matter is found
for $n < 0.02\, n_0$ and consequently is $v_e=v_a$ there.  A
noticeable deviation happens at the onset of the $\Lambda$ hyperon
where $v_e$ drops suddenly, followed by a continuous increase
after the rise of the heavier $\Xi^-$. The same kind of structure
associated with the presence of the hyperons has been observed in
\cite{AGUIRRE,MOTTA}.  The conformal limit
$v_\text{lim}=c/\sqrt{3}$ is clearly exceeded near the $\Lambda$
onset, the maximum value obtained for the speed of sound are
slightly greater than the proposed upper limit $v_\text{max}/c\geq
0.63$ \cite{ALSING}. Before the deconfinement density is $v_e/c
\simeq 0.69$ and a noticeable fall of roughly
$60 \%$ is registered after this particular point.\\
\begin{figure}[t]\vspace{-2cm}\begin{center}
\includegraphics[width=0.8\textwidth]{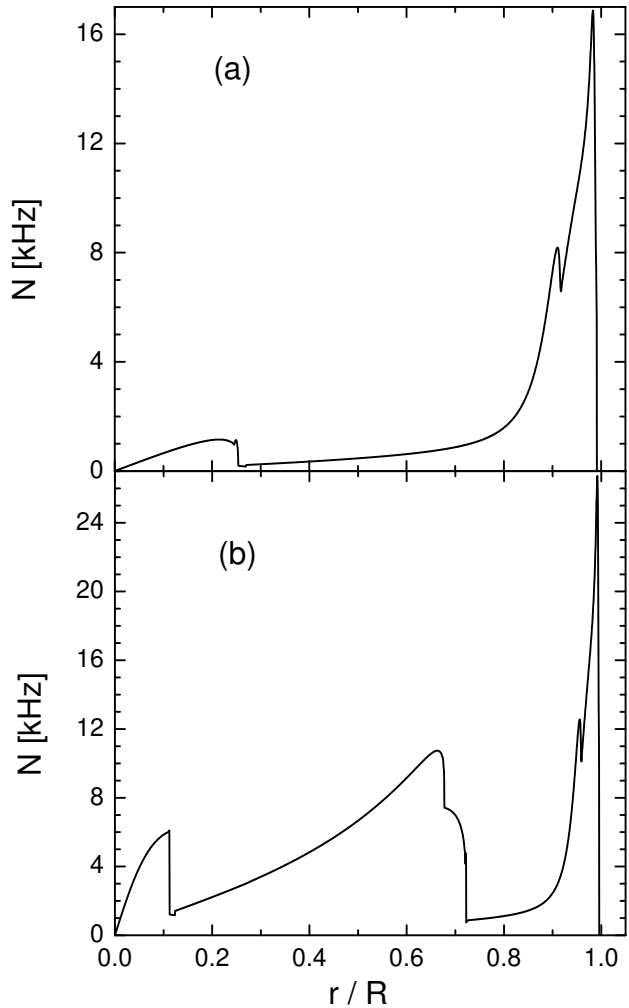}\end{center}
\caption{\footnotesize The Brunt-V\"{a}is\"{a}l\"{a} frequency as
defined by Eq. (\ref{BRUNT}) in terms of the radial coordinate for
a star with $M/M_\odot=1.4$ (a) and  $M/M_\odot=2$ (b), obtained
within the F case. }
\end{figure}
 These observations seems to corroborate the relation
between the magnitude of the speed of sound and the number $N$ of
effective degrees of freedom. In agreement with the general
belief, an increase of $N$ with the density is locally reflected
by a sudden drop in $v_e$, which is realized through a finite
discontinuity in the case of the phase transition. The growth of
$v_e$ observed between these particular points is consistent with
the monotonously increasing trend found in \cite{MOUSTAKIDIS},
where a variety of nuclear matter equations of state are analyzed.\\
Closely related to the sound speed is the relativistic
Brunt-V\"{a}is\"{a}l\"{a} frequency $N$, which is shown in Fig. 8
in terms of the radial coordinate for the stars with
$M/M_\odot=1.4$ (a) and $2.0$ (b).  For the sake of simplicity the
assumption of homogeneity is extended to all the range, which
leads to $N=0$ at the crust of these stars.  The quantity $N$ is a
registry of the bulk properties inside the star, for instance the
onset of the muons at the inner crust causes the small
irregularity mounted on the left side of the peak at $r\simeq R$.
At a deeper point one finds a crest corresponding to the onset of
the hyperons. It takes place at the core of the star in Fig. 8a,
while in Fig. 8b it has a complex structure due to the greater
abundance of $\Lambda$ and $\Xi^-$ for the heavier star. As
already mentioned only a scarce population of $\Lambda$ appears in
the case of $M/M_\odot=1.4$. Finally, a second crest marks the
deconfinement and the coexistence of phases at the center of the
star shown in Fig. 8b, but it does not occur in Fig. 8a.\\
\begin{figure}[t]\vspace{-2cm}\begin{center}
\includegraphics[width=0.8\textwidth]{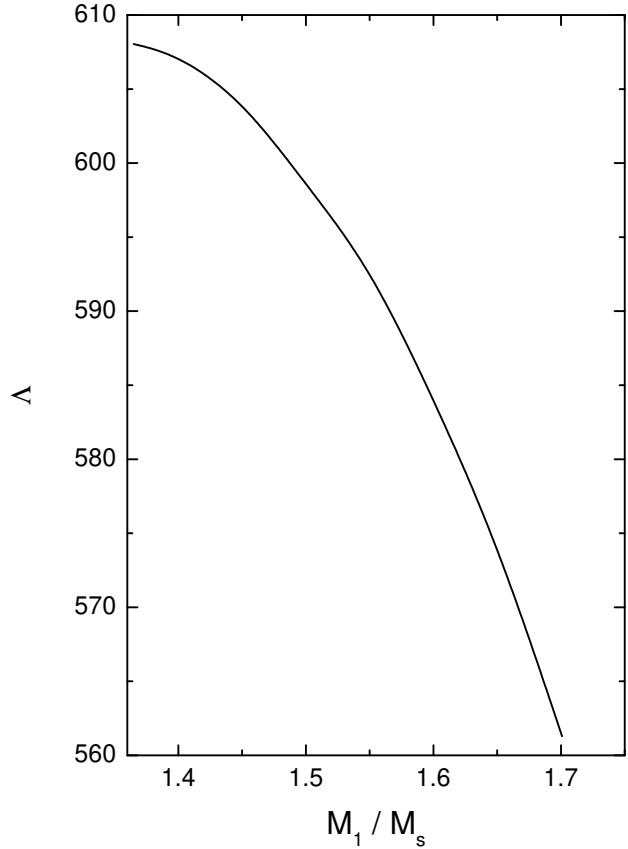}\end{center}
\caption{\footnotesize The combined tidal deformability  of a
binary system (see Eq. (\ref{CTidal})) in terms of the mass of the
heavier component, as obtained in the F case. }
\end{figure}
The possible value for the tidal deformability of  a neutron star
with $M/M_\odot= 1.4$ is within the range
$\Lambda_{1.4}=190^{+390}_{-120}$, according to the analysis of
\cite{ABBOT2}. In the present calculations the result
$\Lambda_{1.4}=527.08$ has been obtained, hence it is compatible
with that constraint. It must be pointed out that the star with
the canonical mass has, in the present analysis, only a $1 \%$ of
$\Lambda$ hyperons in its core. Therefore $\Lambda_{1.4}$ would
not provide information about exotic degrees of freedom.
\\
Another parameter of interest is the combined tidal deformability
\begin{equation}\tilde{\Lambda}=\frac{16}{13}\,\frac{\Lambda_1\,(M_1+12\,M_2)
M_1^4+ \Lambda_2\,(M_2+12\,M_1)
M_2^4}{(M_1+M_2)^5}\label{CTidal}\end{equation}
where $M_i, \,\Lambda_i$ are the mass and the tidal deformability
of the individual components of a binary system. Furthermore the
chirp mass, given by the relation
\[{\cal M}^5=\frac{M_1^3 M_2^3}{M_1+M_2},\]
has been determined with accuracy for the event GW170817
\cite{ABBOT2}, while the possible values for $M_1$ are expected to
range within $1.3 < M_1/M_\odot< 1.6$, assuming  $M_2<M_1$. Under
this constraint I have evaluated $\tilde{\Lambda}$ in terms of
$M_1$, as shown in Fig. 9. The result $560 < \tilde{\Lambda} <
610$ is compatible with the expectations for the low spin prior
$\tilde{\Lambda} \in (70,800)$ as well for the high spin
prior $\tilde{\Lambda} \in (0,630)$ \cite{ABBOT3}.\\
The particular cases of $\Lambda_{1.4}=527.05$ and
$\Lambda_{2.01}=22.2$ have been compared with the predictions of
the universal relation proposed in \cite{YAGI} for
$\ln\left(I/M^3\right)$ in terms of the tidal deformability. In
the first case the discrepancy is negligible and for the more
massive case an agreement within $0.1 \%$ is found.\\
The general compatibility with the main phenomenological data on
the tidal deformability distinguishes the present treatment in
comparison with other models of hyperonic stars \cite{FORTIN}.

It has been argued that the nonradial oscillation modes of a
neutron star can be used to infer structure parameters, such as
mass and radius \cite{ANDERSSON}, or even they can reveal the high
density hadronic EoS \cite{KOKKOTAS} and the presence of exotic
degrees of freedom \cite{SOTANIt}.  The spectrum of non-radial
oscillations for a compact star containing hyperons has been
intensively studied \cite{SOTANIk,BLAZQUEZ,PRADHAN,KUMAR
MISHRA,HYP&GW}. For this reason the fundamental frequency $\nu_f$
of the f-modes, characterized by the fact that the corresponding
radial function $W(r)$  does not have nodes inside the star, and
the highest frequency $\nu_g$ of the g-mode are examined in the
following. The results for these frequencies in terms of the
inertial mass are shown in Fig. 10. Both instances, F and N
treatments, are considered but only stars within the range of
masses allowed by the first case are included. For low $M$ there
is no appreciable difference because such stars have a relatively
low  central density $n_c$ and the EoS are practically the same in
both cases. But as $n_c > 2.5 n_0$ the presence of hyperons
modifies significantly the composition of the core in the F case.
Thus the difference becomes apparent for $M/M_\odot > 1.5$. The
behavior of $\nu_g$ resembles that found in \cite{JAIKUMAR,ZHAO},
although in that work the steep growth of the frequency is driven
by the deconfinement transition which is effective even for low
mass neutron stars. In contrast, I found that only stars with a
mass around $M_\text{max}$ have an admixture of hadrons and free
quarks in their cores. The common feature with
\cite{JAIKUMAR,ZHAO} is a sudden rise associated to the
Brunt-V\"{a}is\"{a}l\"{a} frequency, see Fig. 8, although it
has different physical origin. \\
Recent analysis has estimated the disagreement between the Cowling
approximation and the solution of linearized general relativity
for $\nu_g$ is within $10 \%$ \cite{ZHAO}.\\
 In regard of $\nu_f$, the comparison between the F
and N results is similar to the behavior found in \cite{PRADHAN}
for the $m^\ast/m=0.55$ case. The fitting function given in
\cite{PRADHAN} for the quadrupolar $l=2$ component describes our
results within the $5 \%$ accuracy. Contrary to the expectations
the relation for the real angular frecuency $2 \pi \nu_f$ in terms
of the parameter $\eta=\sqrt{M^3/I}$ presented in \cite{LAU}, does
not describe appropriately our calculations. The cause is twofold:
the sampling of models taken in \cite{LAU} does not include
hyperons, therefore a significant discrepancy is expected for high
$M$. But such cases are precisely the more relevant for the
f-mode, since they include the high density regime. And secondly,
the use of the relativistic Cowling approximation which induces an
error estimated within $30 \%$ \cite{KUNJIPURAYIL}.\\
The g-modes are sustained by the buoyancy restoration force due to
the lack of chemical homogeneity in a stratified structure. Thus
the onset of new degrees of freedom constrained by a chemical
potential enlarges the composition gradients and consequently
reinforces the frequency $\nu_g$. In the F case, such effect could
be multiple, to the presence of hyperons in stars with masses
$M/M_\odot > 1.4$, must be added the onset of deconfined quarks
for those with $M/M_\odot\geq 2$. For this reason the g-modes have
been focused as candidates for revealing the behavior of matter in
the deep interior of a star.\\
\begin{figure}[t]\vspace{-2cm}\begin{center}
\includegraphics[width=0.8\textwidth]{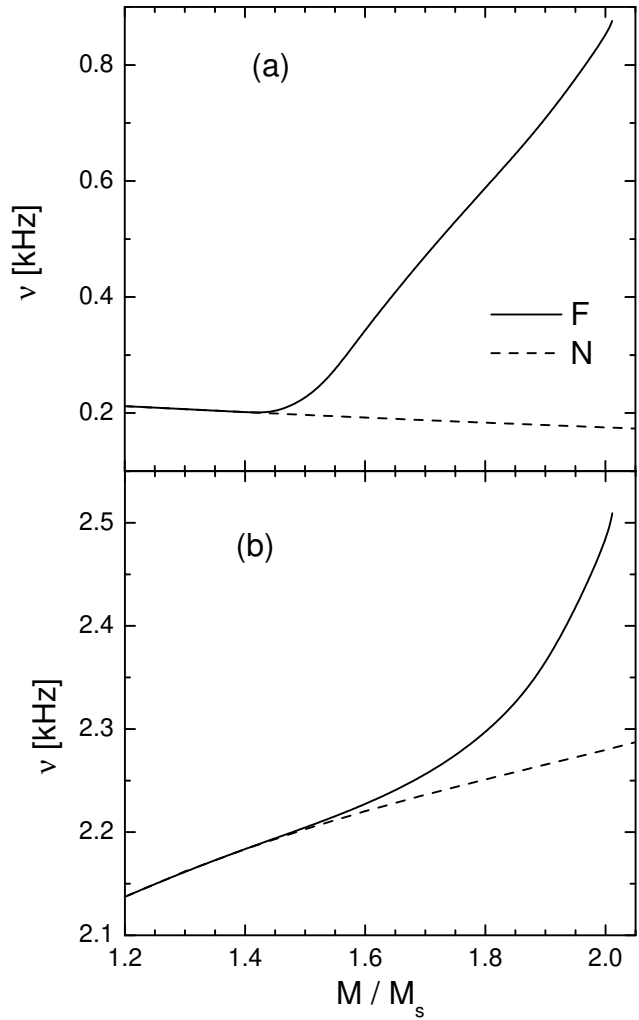}\end{center}
\caption{\footnotesize The fundamental frequency of non-radial
oscillations for the g-mode (a), and the f-mode (b) as functions
of the star mass.  The abbreviations F and N are described in the
caption of Fig. 1.}
\end{figure}
It is expected that the g-modes affect the phase of the
gravitational wave emitted during the collapse of a binary star
system. The underlying mechanism consists in a resonant excitation
of the g-modes by the tidal forces in the last stages of the
inspiral process. Consequently an energy transfer $\Delta E$ from
the orbital motion to the oscillatory dynamics takes place,
modifying the phase of the gravitational waveform. The orbital
phase shift $\Delta \Phi$ due to such mechanism has been estimated
in \cite{REISENEGGER}
\[\Delta \Phi\approx \frac{3}{2}\,\pi\, \Gamma\,\left(\frac{ \Omega_e(t)}{\pi \nu_g}-1\right)
\left(-\frac{\Delta E}{E}\right)\]
and used to study  the effects of hyperons \cite{YU} and quarks
\cite{SALINAS,JAIKUMAR} in the composition of neutron stars. In
the preceding equation $E, \,\Omega_e(t)$ stand for the total
orbital energy and the time dependent orbital angular velocity.
The same analysis is applied here in order to give a comparative
estimation of the effects obtained in the F and N approaches.
Neglecting differences in the time evolution of
$\Omega_e(t)/\nu_g$ the following expression is obtained
\[\frac{\Delta \Phi_F}{\Delta \Phi_N}\simeq \left(\frac{\nu_{g N}}{\nu_{g F}}\right)^{1/3}\,
\left(\frac{S_F}{S_N}\right)^2\]
where the overlap integral $S$ can be evaluated using the
functions $V(r),\,W(r)$ of the  Eqs. \ref{TRANSV}-\ref{RADIAL}.
Taking the case of a star with mass $M/M_\odot=1.8$ the quotients
$\nu_{g F}/\nu_{g N} \simeq 3.5, \, S_F/S_N\simeq 1.5$ are
obtained, hence $\Delta \Phi_F \simeq 1.5\,\Delta \Phi_N$. Taking
as a reference value the result of \cite{JAIKUMAR} $\Delta
\Phi_N\simeq 0.8$, it is found that $\Delta \Phi_F\sim
\mathcal{O}(1)$, which means a noticeable correction to the phase
shift due to the tidal coupling to the oscillation mode. \\
The emission of gravitational waves by a pulsating star at a
distance $R$ apart will be distinguishable by a Earth based
detector if the energy released $E_{GW}$ verifies \cite{KOKKOTAS}
\[\frac{E_{GW}}{M_\odot}=3.5\times 10^{36} \frac{1+4 Q^2}{4 Q^2}\,
S_n[\text{seg}] \left(\frac{S}{N}\right)^2 R[10\, \text{kpc}]^2
\,\nu[\text{Khz}]^2 \]
where $S/N$ is the signal to noise ratio, $S_n$ the spectral
density of the detector, and the quality factor of the oscillation
is given by $Q=\pi\,\nu\,\tau$. The damping time of the
oscillation $\tau$ has not been evaluated here, for these reason
the range $10^{-2}\, \text{year}\leq \tau \leq 3 \times 10^{-2}\,
\text{year}$ is considered according to typical values obtained in
\cite{SALINAS, ZHENG} for a star with $M/M_\odot=1.8$. Adopting
the reference values $S/N=8,\,S_n=4\times 10^{-46} \text{s}$
\cite{ZHENG} corresponding to present LIGO/Virgo characteristics,
the results $E_{GW}^{(N)}\simeq 0.7\times 10^{46}$ erg, and
$E_{GW}^{(F)}\simeq 8 \times 10^{46}$ are obtained for $R=10$ kpc.
Whereas for $R=10$ Mpc the outcomes are $E_{GW}^{(N)}\simeq
1.6\times 10^{52}$ erg, and $E_{GW}^{(F)}\simeq 2 \times 10^{53}$.
Since the LIGO/Virgo energy thresholds for detection is about
$10^{46}-10^{47}$ erg for events in our galaxy, and
$10^{52}-10^{53}$ erg for sources in the Virgo cluster, it can be
concluded that such illustrative situation is feasible of
detection. Furthermore, the relative error in the determination of
the frequency has been estimated as \cite{KOKKOTAS}
\[\frac{\Delta \nu}{\nu}=4.2\times 10^{-3}\,\frac{R[10\,\text{kpc}]}{\tau [\text{s}]}\,
 \sqrt{\frac{1-2 Q^2+8 Q^4}{4 Q^4}}\,\sqrt{\frac{S_n[10^{-46} \text{Hz}]}{E_{GW}/M_\odot}}
\]
Using this relation, an indetermination of only a few Hz is
expected for both N and F cases. Therefore they could be perfectly
discernible.\\
Of course this is a simplified qualitative estimation, a detailed
calculation must take into account the statistical error coming
from other several parameters of the binary that could affect
$\Delta \Phi$ as well $\Delta \nu$.

There is a general belief that hadronic matter at sufficiently
high density undergoes a transition to deconfined quark matter.
However, the characteristics of such transition are uncertain yet.
In this work it is assumed that a preliminary coexistence of
phases takes place, which has been  interpreted as the consequence
of a vanishing interface tension $\sigma_T$. At the opposite
extreme, for very large $\sigma_T$, it is expected a discontinuous
transition described by the Maxwell construction. While for
intermediate values a non-homogeneous phase would be plausible.
These effects have been analyzed in \cite{XIA} within a specific
model, concluding that all of them, the maximum mass, the radius,
and the combined tidal deformability monotonously increase with
$\sigma_T$. An estimation of the maximum variation due to finite
tension is given there as $\Delta M_\text{max}/M_\odot=0.02,\,
\Delta R=0.6$ km, and $\Delta \tilde{\Lambda}/\tilde{\Lambda}=0.5$
\cite{XIA}. Thus a scarce increase in the maximum mass can be
obtained at the cost of a small growth of the radius and a
considerable increment of the tidal deformability.\\

The details of the composition of extreme density matter are still
speculative, although they can be a determining factor for the
structure of compact stars. Different hypothesis has been
explored, as for instance superconducting quark matter, giving
rise to a variety of effective theoretical models. This
uncertainty is evidenced by the amplitude of values assigned to
certain model parameters such as the quark-quark coupling constant
or the energy gap between normal and paired states.
 A large number of studies have discussed the effects of
superconducting quark matter have on the properties of compact
stars \cite{ALFORD2,BUBALLA,LAWLEY,PAGLIARA,PAULUCCI}. For
instance in \cite{ALFORD2} an effective nuclear model is used in
combination with a bag model including a color-flavor locked
superconducting phase. For the latter model the parameters are
taken as $B=137$ MeV/fm$^3$, $m_s=200$ MeV and $\Delta=100$ MeV
for the energy gap. The mass-radius relation for the neutron star
shows the significant fact that a sharp quark-hadron phase
transition leads to a unstable star structure. In contrast, the
continuous phase transition allows the existence of stable
configurations with unbound quarks. In any case the maximum mass
is slightly reduced as compared with the unpaired case. This
behavior is qualitatively confirmed in \cite{BUBALLA} where the
dynamics of the deconfined quarks is determined by the NJL within
two different parametrizations. Since the quark-quark interaction
is unknown, the authors assume the same coupling constant as in
the four fields quark-antiquark interaction $G_D=G$. They only
consider a sharp hadron-quark phase transition and also include
the possibility of light quark superconducting phase (2SC), with
unpaired strange flavor, in addition to the just mentioned
color-flavor locked arrangement. In this case, the presence of the
intermediate two-flavor pairing introduces a narrow window of
stability before the color-flavor locked phase becomes preferable.\\
These type of instabilities have been related to the lack of
confinement of the NJL model \cite{BALDO}, and attributed to the
zero energy point $\mathcal{E}_0$. This argument has been examined
in \cite{PAGLIARA}, where a different procedure to fix the
additive constant  has been proposed. With this modified constant
$\mathcal{E}_0^*$, an intermediate stable 2SC phase was found, as
in \cite{BUBALLA}. A further increase of the pairing coupling
constant to $G_D=1.2 \,G$, in combination with $\mathcal{E}_0^*$,
extends the range of stability to embrace the color-flavor locked
phase. At the same time the allowed maximum mass for neutron stars
is reduced \cite{PAGLIARA}. \\Based on this results one can
conclude that the inclusion of a superconducting quark phase, if
stable, will lead to a decrease of $M_\text{ max}$.

\section{Summary and Conclusions}\label{Sec5}

This work is devoted to the study of dense matter at zero
temperature, as can be found in the interior of neutron stars. For
this purpose a composite model of the strong interaction is used.
In the low density limit a nonhomogeneous phase including light
atomic nuclei is considered through the standard results of
\cite{BPS}. For higher densities a  model of the field theory
including hyperons is used to describe a homogeneous hadronic
phase. For the extreme densities feasible in the core of a compact
star, a phase of deconfined quarks is taken into account through
the NJL model with vector current-current coupling. In between a
coexistence of hadronic and free quark phases is assumed, which
allows a continuous variation of the
thermodynamic potential.\\
The role of hyperons is particularly analyzed in the context of
the ``hyperon puzzle". Thus an extension of a previously defined
hadronic model \cite{MCS,AGUIRRE} is made by including
hyperon-hyperon interaction mediated by the $a_0(980)$ and hidden
strangeness $f_0(980)$ mesons. In order to emphasize the hyperonic
effects a complementary scheme, which only considers nucleons and
leptons, is introduced.\\
Several properties of a static or slowly rotating neutron star has
been evaluated, such as maximum mass, moment of inertia, tidal
deformability, etc., and contrasted with recent observational data
or with different universal relations. Since the nature of the
compact object with mass $M/M_\odot \simeq 2.6$ detected by
\cite{ABBOTT0} is still uncertain, it is not actively considered
here. The confirmation that it is a neutron star would put most of
the present models of the strong interaction satisfying explicit
relativistic covariance and agreement with other observational
data in conflict.\\
Within this approach only stars with masses $M/M_\odot\geq 1.45$
show clear evidence of the presence of hyperons, and only those
with $M/M_\odot\simeq 2$ have a trace of deconfined quarks in
their cores.\\
The pressure at a density twice the normal nuclear density is
coherent with the estimations based on the data obtained in the
GW170817 event \cite{ABBOT2}.  The mass-radius relation for the
maximum mass predicted in the present treatment $M/M_\odot= 2.01,
\; R=11.7$ km is within the confidence range proposed for the PSR
J0740+6620 \cite{RILEY2}.\\
The imminent high precision measurement of the moment of inertia
of the PSR J0737-3039A has motivated intense work. Focusing on a
star with the same mass, our result for the moment of inertia is
compatible with different predictions \cite{LIM,LANDRY}.\\
Focusing on a star with the canonical mass $M/M_\odot=1.4$, I
found its radius is in agreement with the analysis in
\cite{RAAIJMAKERS,MILLER}. Furthermore, its tidal deformability
$\Lambda_{1.4}$ verifies the constraint $\Lambda_{1/4} < 580$
established in \cite{ABBOT2}. Considering a binary system with
chirp mass ${\cal M}/M_\odot=1.186$ the result for the combined
tidal deformability verifies $566 < \tilde{\Lambda} < 608$ which
is compatible with the observational evidence \cite{ABBOT3}.\\
The calculations for the frequencies $\nu$ of the nonradial f and
g-modes for a neutron star are qualitatively similar to previous
results for either including hyperons (F) or considering only
nucleons and leptons (N). The Brunt-V\"{a}is\"{a}l\"{a} frequency
have peaks associated with the onset of the hyperons similar to
those found for the deconfinement transition \cite{JAIKUMAR}. They
are the cause of important deviations of the frequencies found in
the F case as compared with the N treatment. This is a
confirmation of previous findings for $\nu_g$ \cite{JAIKUMAR} as
well as for $\nu_f$ \cite{PRADHAN}.\\
In addition, positive confirmations of our results are obtained
when contrasting with the universal relations proposed for the
moment of inertia in terms of the compactness \cite{LIM,LATTIMER},
the tidal deformability as a function of the moment of inertia
\cite{YAGI}, and the f-mode frequency in terms of $M/R^3$
\cite{PRADHAN}.

In summary, the model used here requires only a concise set of
parameters inspired on basic phenomenological grounds and its
predictions are in good agreement with a variety of recent
observational data as well as phenomenological relations on
neutron stars. Hence one can conclude that the presence of
hyperons and a deconfinement transition are compatible with the
present knowledge on compact stars.\\

\section*{Acknowledgements}This work was partially supported by the
CONICET, Argentina under grant PIP-616.

\end{document}